\documentclass[conference, twocolumn]{IEEEtran}
\normalsize
\ifCLASSINFOpdf
\else
\fi

\usepackage{amsthm}
\usepackage{amsmath}
\usepackage{amssymb}
\usepackage{graphicx}
\usepackage{esint}
\usepackage{amsfonts}
\usepackage{cite}
\usepackage{balance}
\usepackage{caption}
\usepackage{subcaption}
\usepackage{epstopdf}
\usepackage{color}
\makeatletter
\usepackage{amssymb}
\usepackage{gensymb}
\usepackage{lipsum}% http://ctan.org/pkg/lipsum
\usepackage{algorithm}% http://ctan.org/pkg/algorithm
\usepackage{algpseudocode}% http://ctan.org/pkg/algorithmicx
\usepackage[compatibility=false]{caption}% http://ctan.org/pkg/caption
\usepackage[bottom]{footmisc}
\theoremstyle{plain}

\makeatother

\providecommand{\theoremname}{Theorem}

\theoremstyle{plain}

\makeatother

\providecommand{\lemmaname}{Lemma}
\DeclareMathOperator{\arctantwo}{arctan2}

\newcounter{a}

\begin{document}
\title{A cooperative localization-aided tracking algorithm for THz wireless  systems}

\author{%

\authorblockN{Giorgos Stratidakis, Alexandros--Apostolos A. Boulogeorgos\authorrefmark{1}, Angeliki~Alexiou\authorrefmark{1}}

\authorblockA{\authorrefmark{1}\footnotesize  Department of Digital Systems, University of Piraeus, Piraeus 18534,

Greece. (E-mails: al.boulogeorgos@ieee.org, alexiou@unipi.gr).}

}
\author{

Giorgos Stratidakis, Alexandros--Apostolos A. Boulogeorgos, and~Angeliki~Alexiou
\\

\begin{normalsize} 

Department of Digital Systems, University of Piraeus, Piraeus 18534, Greece.

\end{normalsize}

\\

\begin{normalsize} 

e-mail: giostrat@unipi.gr, al.boulogeorgos@ieee.org, alexiou@unipi.gr.

\end{normalsize} 

}

\maketitle	

\begin{abstract}
In this paper, a novel cooperation-aided localization and tracking approach, suitable for terahertz (THz) wireless systems is presented. It combines an angle of arrival (AoA) tracking algorithm with the two-way time of arrival method, in order to accurately track the user equipments (UE) position and reduce the deafness caused by the estimation errors of the tracking algorithms. This algorithm can be used by one base station (BS) to estimate the UEs position, or by multiple BSs, that cooperate with each other to increase the accuracy of the estimations, as well as the probability of successful estimations and guarantee low-estimation overhead. The efficiency of the algorithm is evaluated in terms of deafness and probability of successful AoA estimation and is compared with the corresponding performance of the fast channel tracking algorithm.

\end{abstract}


\section{Introduction}\label{S:Intro}
In recent years, there has been an important increase in wireless services with a corresponding need for data rates~\cite{PhD:Boulogeorgos,A:LC_CR_vs_SS}. Terahertz (THz) communications promise to fill the data rate demand by using the huge amount of available non-standardized bandwidth in frequencies from $0.1$ to $10$ THz~\cite{A:Analytical_Performance_Assessment_of_THz_Wireless_Systems}. On the other hand, communications in these frequencies suffer from huge channel attenuations \cite{Jornet2011,our_PIMRC,Kokkoniemi2018,C:PerfEvaluation,our_spawc_paper_2018,C:UserAssociationInUltraDenseTHzNetworks}. To account for the increased losses, THz systems employ large antenna arrays to form pencil-beams, with high antenna gain \cite{Boulogeorgos2018,WP:Wireless_Thz_system_architecture_for_networks_beyond_5G,A:Analytical_Performance_Assessment_of_THz_Wireless_Systems}. In order for pencil-beamforming to work, the base station must know and track the direction of the user equipment (UE), or the connection will suffer from deafness. The localization and tracking techniques so far, require a significant increase in overhead with the reduction of beamwidth. As a result, they cannot be used efficiently in THz systems. Therefore, the development of localization techniques in THz systems with low overhead is an important task.\\
\indent Scanning the open literature, there are several published contributions that report localization and tracking algorithms (see for example \cite{Gao2017,Zhang2016,Va2016} and references therein). In more detail, in \cite{Gao2017} the authors proposed a prediction algorithm, in order to track the direction of the UE, which allowed accurate beam-tracking with low overhead. Unfortunately, if the UE's does not follow a linear motion, the prediction may fail and the tracking has to start again, which means an increase in overhead. Moreover, the estimation is based on the strength of the elements of the beamspace channel, which means that the directions that can be estimated are specific, regardless of the actual directions and results in power leakage. In \cite{Zhang2016}, a Kalman-filter based algorithm and an abrupt change detection were employed to track the UE. However, the overhead, which was required in order to guarantee an accurate UE tracking, dramatically increases with the number of antennas. In other words, this approach would require an extremely high overhead in THz pencil-beamforming systems. Likewise, in \cite{Va2016}, the proposed algorithm employed an extended Kalman filter in order to employ only one measurement of a single beam-pair to track the propagation path, but assumed that the devices can change the antenna pattern to any arbitrary direction, which in practice is infeasible \cite{Jayaprakasam2017}.\\
\indent All the above mentioned works have taken into account only the physical direction of the UE and neglect its position. Several localization algorithms have been proposed (see for example \cite{Dargie2010} and references therein). In \cite{Dargie2010}, two ToA methods are described, the one-way ToA and the two-way ToA. The one-way ToA needs only 1 message to estimate the distance but needs accurate synchronization of the clocks between the transmitter and the receiver. The two-way ToA needs two messages to estimate the distance but does not need as high accuracy in the synchronization as the one-way ToA. The time difference of arrival (TDoA) method does not require sunchronized clocks, but it needs additional equipment, in order to send two signals with different velocities (e.g., a radio and an acoustic signal).
Moreover, in \cite{Rong2006}, a triangulation approach is presented. However, this approach demands at least three base stations (BSs), that can exchange their AoA estimations.
%Likewise, in \cite{Chen2014} and \cite{Christoph2013}, the received signal strength (RSS) estimation approach was used, which enabled the evaluation of the distance. However, in THz systems, the signal strength cannot be used as a measure, due to the pathloss model needing environmental variables, such as relative humidity and temperature \cite{Kokkoniemi2018}. 
On the other hand, global positioning system (GPS), is widely used in outdoor localization but is not available in indoor environments as the sattelite signal is not available in most cases \cite{Drawil2013}.
Unfortunately, both triangulation and trilateration need at least three measurements to estimate a unique location, which may not be the case in THz pencil beamforming systems due to non-line of sight (NLOS).
Cooperative localization algorithms, have attracted a considerable amount of interest in THz systems, as they increase the performance of localization in both accuracy and coverage \cite{Wymeersch2009,Ouyang2010}. In more detail, in \cite{Wymeersch2009,Ouyang2010}, the authors employ AoA, ToA and TDoA methods for an anchor to locate the desired node and exchange their estimations with the other anchors.\\
\indent Motivated by the above, in this paper, a novel cooperative localization approach, suitable for THz wireless systems with pencil-beamforming, which can be used for non-linear motions tracking,  is presented. It employs an angle of arrival (AoA) tracking algorithm and the two-way time of arrival method, in order to track the UE’s position with only one base station (BS). Furthermore, multiple BSs are used, which cooperate with each other, in order to combine their estimations and increase the localization accuracy, while guaranteeing  low-estimation overhead. The efficiency of the algorithm is evaluated in terms of deafness and probability of successful AoA estimation and is compared with the corresponding performance of the fast channel tracking (FTC) algorithm.
\subsection{Notations} 
Unless otherwise stated, lower case and upper case bold letters denote a vector and a matrix, respectively; $\mathbf{A}^H$ denotes the conjugate transpose, $\mathbf{A}^{-1}$ denotes the inversion, and $tr(\mathbf{A})$ denotes the trace of matrix $\mathbf{A}$; $|a|$ denotes the amplitude of scalar a; card($\mathbf{A}$) denotes the cardinality of set $\mathbf{A}$; $\mathrm{supp}(\mathbf{A})$ denotes the support of set $\mathbf{A}$; $\bmod_N(\left(\cdot\right))$ denotes is the modulo operation with respect to $N$; $\arg\min(\mathbf{A})$ denotes the index of set $\mathbf{A}$ at which the values of $\mathbf{A}$ are minimized; and finally, $\mathbf{I_{K}}$ is the $K \times K$ identity matrix.

 %It uses the fast channel tracking (FCT) algorithm in \cite{Gao2017}, which is a beamspace method to estimate the AoA, due to the low pilot overhead and the two-way ToA to estimate the distance due to the low clock synchronization without additional equipment to estimate the distance. The combination of these two makes localization relatively easy. Moreover, it enables localization by one BS. As a result, the sharing of the UEs position, with other BSs becomes available. The shared information of the positions can be combined to increase the accuracy of the position estimations, increase the probability of successful AoA estimations and fix the fundamental weakness of the beamspace methods, which prevents them from tracking all angles. Furthermore, the localization enables the tracking of the UE, without communication (blind tracking), for the BSs that did not participate in the estimations and guarantees the low overhead that the FCT algorithm promises. 

\section{System and signal model}\label{sec:SSM}
\begin{figure}
\centering
%\captionsetup{justification=centering}
\includegraphics[width=1\linewidth,trim=0 0 0 0,clip=false]{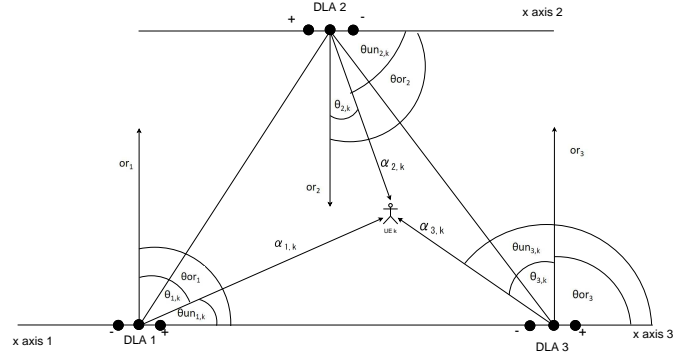}
\caption{An example of the system model under consideration, with 3 BSs and 1 UE}
\label{fig:angl_dist}
\end{figure}

An indoor THz system, in which three BSs are used to serve $K$ UEs is assumed. Each BS is equipped with a single-sided discrete lens array (DLA)\footnote{Discrete lens arrays (DLAs) have been employed in milimeter wave (mmWave) and THz communication as a low-energy consumption MIMO alternatives.} that employs $N$ elements and $N_{\mathrm{RF}}\leq N$ radio frequency (RF) front-end chains. Each BS receives a signal with different channel and from a different direction (relative to their position) from the others. As illustrated in Fig. \ref{fig:angl_dist}, a two dimensional cartesian plane is considered, without obstacles between the BSs and the UEs. In this figure, $\theta_{or_{i}}$ is the angle between the x-axis of each BS in their individual coordinate system and the orientation $or_i$ of the $i$-th DLA, $\theta_{i,k}$ is the angle between the orientation of the $i$-th BS and the $k$-th UE, $\alpha_{i,k}$ is the distance between the $i$-th BS and the $k$-th UE and $\theta_{un_{i,k}}$ is the angle between the positive positive (if $\theta_{or_i}$ is positive, and negative if $\theta_{or_i}$ is negative) x-axis\footnote{The x-axis of all the BSs are parallel to each other.} of the $i$-th BS and the ray to the $k$-th UE. The orientation is defined as a fixed direction against, which the AoAs are measured \cite{Rong2006}. If the UE is on the negative side, in relation to the BS, then $\theta_{i,k}$ is negative and if it is on the positive side, it is positive. Furthermore, it is assumed that the BSs communicate with a common node, which plays the role of the fusion center and collects the information about the predicted position of the UE by each BS. The fusion center can be either a new node or a predefined BS. For simplicity and without loss of generality, it is assumed that $K=N_{\mathrm{RF}}$. Hence, the baseband equivalent received signal vector for the $i$-th BS and the $k$-th UE can be obtained as

\begin{align}
%y_{i,k}=H_{i,k}^H ~F ~x+~n= H_{i,k}^H ~F ~Ps+~n,
%\mathbf{y}_{i,k}= \tilde{\mathbf{H}}_{i,k}^H  ~\mathbf{x}+~\mathbf{z}_i= \mathbf{U}\mathbf{H}_{i,k}^H ~\mathbf{Ps}+~\mathbf{z}_i,
\tilde{\mathbf{y}_{i}}= \tilde{\mathbf{H}}_{i}^H  ~\mathbf{x}+~\mathbf{z}_i= \mathbf{H}_{i}^H \mathbf{U}^H ~\mathbf{Ps}+~\mathbf{z}_i,
\label{Eq:Received_signal}
\end{align}
where $\tilde{\mathbf{H}}_{i}=[\tilde{\mathbf{h}}_{i,1},\tilde{\mathbf{h}}_{i,2},…,\tilde{\mathbf{h}}_{i,k}]$ is the beamspace channel matrix between the $i$-th BS and the $k$-th UE, $\mathbf{\tilde{\mathbf{h}}}_{i,k}$ is the beamspace channel vector between the $i$-th BS and the $k$-th UE, $\mathbf{U}$ is the discrete Fourier transform (DFT) matrix that corresponds to the DLA \cite{Sayeed2013}, $\mathbf{H}_{i}=[\mathbf{h}_{i,1},\mathbf{h}_{i,2},…,\mathbf{h}_{i,k}]$ is the MIMO channel matrix between the $i$-th BS and the $k$-th UE~and 
\begin{align}
\mathbf{x}= \mathbf{Ps}
\label{Eq:Transm_sig_vec}
\end{align}
is the transmitted signal vector. In \ref{Eq:Transm_sig_vec}, $\mathbf{s}$ stands for the original transmitted signal vector for all $K$ UEs, with normalized power $E(\mathbf{ss}^H)=\mathbf{I}_K$, with $\mathbf{I}_K$ being the identity matrix and $\mathbf{P}$ is the precoding matrix satisfying the total transmit power constraint as $\mathrm{tr}(\mathbf{PP}^H )\le \rho $, where $\rho$ is the total transmit power. Moreover, $\mathbf{z}_i$ is the additive Gaussian noise (AWGN) vector of the $i$-th BS. Finaly, the matrix $\mathbf{U}$ consists of the array steering vectors of N orthogonal directions that cover the entire angular domain and can be obtained as
\begin{align}
\mathbf{U} = [\mathbf{a}(\tilde{\psi}_1),\mathbf{a}(\tilde{\psi}_2),...,\mathbf{a}(\tilde{\psi}_N)]^H,
\label{Eq:DFT_matrix}
\end{align}
where 
\begin{align}
\tilde{\psi}_n= \dfrac{1}{N}\left(n-\dfrac{N+1}{2}\right),
\label{Eq:norm_sp_dir}
\end{align}
 with $n=1,2,...,N $ being the normalized spatial directions, which are predefined by the DLA. For the shake of convenience, it is assumed that the channel state information (CSI) at the BSs is perfect.\\
\indent Next, we present the channel model. In this paper, we employ the Saleh-Valenzuela channel model as \cite{Gao2017}, \cite{Sayeed2013} 
\begin{align}
\mathbf{h}_{i,k} = \beta_{k}^{(0)} \mathbf{a}(\psi_k^{(0)})+ \sum_{i=1}^{L} \beta_k^{(i)} \mathbf{a}(\psi_k^{(i)}),
%h_{(\alpha,\theta)} = \overbrace{\dfrac{a}{\sqrt{N_b N_m}}(\sum_{b=1}^{Nb} \sum_{m=1}^{Nm} w_b^* f_m e^{jkd(b-m)\cos{\theta}})q}^{h(\alpha,\theta)},
\label{Eq:ch_vec}
\end{align}
where $\beta_{k}^{(0)} \mathbf{a}(\psi_k^{(0)})$ is the line of sight (LoS) component of the $k$-th, $\sum_{i=1}^{L} \beta_k^{(i)} \mathbf{a}(\psi_k^{(i)})$ is the NLoS component and $L$ is the number of NLoS components. Furthermore, $\beta_{k}^{(0)}$ and $\beta_k^{(i)}$ are the complex gains, while $\psi_k^{(0)}$ and $\psi_k^{(i)}$ represent the spatial directions. Note, that the NLoS components are typically weaker than the LoS component due to scattering. In THz frequencies, scattering induces more than 20 dB attenuation in the NLoS components \cite{our_PIMRC}. As a result, only the LoS component can be used reliably in THz systems. Therefore, the MIMO channel vector $h_{k}$ can be simplified as

\begin{align}
\mathbf{h}_{i,k} = \beta_{k} \mathbf{a}(\psi_k).
\label{Eq:ch_vec2}
\end{align}

In a typical uniform linear array (ULA) with $N$ antennas, the array steering vector can be obtained as \cite{Gao2017}

\begin{align}
\mathbf{a}(\psi)= \dfrac{1}{\sqrt{N}} [e^{-j2\pi \psi m}]_{m\in \mathcal{I}(N)},
\label{Eq:sp_dir}
\end{align}
where $\mathcal{I}(N)={l-(N-1)/2, l=0,1,...,N-1}$ is a symmetric set of indices centered around zero. The spatial direction can be obtained as 
\begin{align}
\psi \triangleq \dfrac{d}{\lambda} \sin \theta,
\end{align}
where $\theta$ is the physical direction, $\lambda$ is the signal wavelength and $d$ is the antenna spacing that usually satisfies $d= \lambda/2$. 

\section{Cooperation-aided localization approach}
The proposed localization method consists of four phases, namely i) Fast Channel Tracking, ii) Ranging, iii) Localization and iv) Cooperation. 
In the first phase, an extention of the tracking algorithm, which was initially proposed in \cite{Gao2017}, is provided to obtain the UE's AoA estimation. 
In the second phase, a two-way ToA approach is used in each BS to obtain its distance from the UE. 
In the third phase, we combine the physical direction of the UE with the distance to extract the UE's position in each BS cartesian coordination system. In the fourth phase, the estimated positions of the UE from all the BSs are sent to a predetermined BS, which converts them into a common coordination system and combines them in order to increase the accuracy of the estimation. Next, a detailed description of the phases is provided as well as Algorithm 1 that illustrates the localisation procedure. In Algorithm 1, $E_r$ and  $\tilde{E}_r$ respectively stand for the estimated and expected energy.

\subsection{Phase 1- Proposed Fast Channel Tracking Algorithm}
The $i$-th BS, in the first three timeslots, estimates the beamspace channel, using conventional beamspace channel estimation schemes, and obtains the strongest element of the beamspace channel, $n_{i,k}$. Then, the AoA can be approximated as
\begin{align}
\mathbf{\theta}_{i,k}(t)\approx \arcsin \dfrac{\lambda}{N d}\left(n_{i,k}(t)- \dfrac{N+1}{2}\right),
\label{Eq:theta_approx}
\end{align}
where $t$ denotes the timeslot index and $\lambda$ is the wavelength. Note, that by using the beamspace channel, to estimate the direction based on $n_{i,k}$, indicates that the estimations are specific and equal to the number of elements of the beamspace channel \cite{Lin2017}. As a result, the algorithm will return one of these directions instead of the actual direction that the UE is at, which causes the estimation error of the direction to be random.

After the first 3 timeslots, the $i$-th BS starts predicting the next AoA of the UE, by using the previous estimations. The localization part of the algorithm enables us to predict the next position of the UE, by assuming a linear motion, as 
\begin{gather}
	\begin{split}
	&\mathbf{\overline{r}}_k(t+1)= \mathbf{\overline{r}}_k(t)+ \\
	&\dfrac{\ [\mathbf{\overline{r}}_k(t)-\mathbf{\overline{r}}_k(t-1)]+[\mathbf{\overline{r}}_k(t-1)- \mathbf{\overline{r}}_k(t-2)]}{2},
	\end{split}
	\label{Eq:pos_pred}
\end{gather}
where $\mathbf{\overline{r}}_k= (\overline{x}_k(t), \overline{y}_k(t))$ is the position of the $k$-th UE estimated in Phase 4, with $\overline{x}_k(t)$ and $\overline{y}_k(t)$ being the coordinates of the UE in the common coordination system.  The AoA of the predicted position can be estimated as 
\begin{gather}
	\begin{split}
	\mathbf{\theta}_{i,k}(t+1)= \mathbf{\theta}_{or_{i}}- \arctantwo(\mathbf{\overline{r}}_k(t+1)-\mathbf{p_i})
	\end{split}
	\label{Eq:theta_pred}
\end{gather}
where the operator $\arctantwo\left(\cdot\right)$ returns the angle between the positive x-axis of the BS and the ray to the UE.

After predicting the next AoA of the UE, the position of the strongest element $n_{i,k}$ of $\tilde{\mathbf{h}}_{i,k}(t)$ can be presented as
\begin{align}
n_{i,k}=\underset{1 \leqslant n \leqslant N}{\arg\min} \Big|\tilde{\psi}_n- \psi_k \Big|= \underset{1 \leqslant n \leqslant N}{\arg\min} \Big|\tilde{\psi}_n- \dfrac{d}{\lambda}\sin(\theta_{i,k}) \Big|,
\label{Eq:n_ik_pre}
\end{align}
The support, i.e., the set of indices of nonzero elements in a sparse vector, of $\tilde{h}_k$ can be determined by $n_{i,k}$, without channel estimation as \cite{Gao2017}
\begin{align}
\mathrm{supp}(\tilde{\mathbf{h}}_{i,k})= \bmod_N \bigg\{{n_{i,k}-\dfrac{V}{2},...,n_{i,k}+\dfrac{V-2}{2}} \bigg\},
\label{Eq:supp}
\end{align}
if $V$ is even and as
\begin{align}
\mathrm{supp}(\tilde{\mathbf{h}}_{i,k})= \bmod_N \bigg\{{n_{i,k}-\dfrac{V-1}{2},...,n_{i,k}+\dfrac{V-1}{2}} \bigg\},
\label{Eq:supp2}
\end{align}
if $V$ is odd, where 
\begin{align}
V= \mathrm{card}(\mathrm{supp}(\mathbf{\tilde{\mathbf{h}}}_{i,k})),
\label{Eq:V_spars}
\end{align}
and $\mathrm{card} \left(\cdot\right)$ denotes the cardinality.

\subsection{Phase 2- Ranging}
In order to estimate the distance between the BS and the UE, a ranging technique is required. The BS sends a message in the direction that was estimated in Phase 1. If the UE is in that direction, it responds, otherwise Phase 1 has to start over. The BS keeps track of the transmitting and receiving time instants of both the transmitted and received messages, and the UE sends its transmitting and receiving time instants as feedback to the BS. The range estimation errors will be corrected in Phase 4. It is assumed that the synchronization between the BS and the UE is perfect. The distance can be calculated as
%Since phase 1 sends pilots in different directions, they can be used to estimate the distance, instead of sending a new message. 

\begin{align}
\alpha_{i,k}=\dfrac{\ (t^4_{i,k}-t^1_{i,k})-(t^3_{i,k}-t^2_{i,k})} {2} c,
\label{Eq:dist}
\end{align}
where $c$ is the speed of light, $t^1_{i,k}$ and $t^2_{i,k}$ are the transmitting and receiving times of the transmit signal of the $i$-th BS to the $k$-th UE and $^t3_{i,k}$ and $t^4_{i,k}$ are the transmitting and receiving times of the response signal of the $k$-th UE to the $i$-th BS.

\subsection{Phase 3- Localization}
After obtaining the angle and the distance of the UE from each BS, we calculate the UE's position. First, the AoA must be converted to the angle between the positive x-axis of the BS and the ray to the UE, in order to calculate the location of UE. The physical direction in the common coordination system can be evaluated as

\begin{align}
\theta_{un_{i,k}}= \theta_{or_{i}}- \theta_{i,k},
\label{eq:th_uni}
\end{align}

The position can then be calculated as \cite{Bronshtein2015}
\begin{align}
\mathbf{r}_{i,k}(t)=\alpha_{i,k}(t)[\cos(\theta_{un_{i,k}}(t)),\sin(\theta_{un_{i,k}}(t))]+ \mathbf{p}_i,
\label{eq:pos}
\end{align}
where $\mathbf{r}_{i,k}$ is the calculated position of the $k$-th UE by the $i$-th BS and $\mathbf{p}_i$ are the coordinates of the BSs. 

\subsection{Phase 4- Cooperation}
After the BSs estimate the position of the UE, they send the information to a predetermined node to refine it by calculating the center of gravity of all the estimated positions as \cite{Chang2018}, \cite{Karl2005}
\begin{align}
\mathbf{\overline{r}}_k = \dfrac{\sum_{i=1}^{M} \mathbf{r}_{i,k}}{M},
\label{Eq:pos_mean}
\end{align}
where $M$ is the number of BSs that estimated the position of the UE.
Using the center of gravity to make a common estimation decreases the overall misalignment. However, when more than one BSs have estimated the position of the UE the misalignment of the most accurate BS is decreased. After calculating the center of gravity, the common node sends the information to all the BSs. As long as one BS estimates the position of the UE, all the other BSs can calculate the $\theta_{i,k}$ of the UE, keep predicting the next $\theta_{k}$ of the UE, and use the low-pilot overhead of FCT. The knowledge of the UE's position, makes the calculation of the AoA relatively simple for all the BSs regardless of their position and the presence of obstacles within the LoS path. Furthermore, as all the BSs in the surrounding area know the UE's position and direction of motion, they can prepare for handover if needed.
The AoA is calculated as
\begin{align}
\theta_{k}(t)= \theta_{or_{i}}- \mathrm{\arctantwo}(\mathbf{\overline{r}}_k-p_i)
\label{eq:th_af_pos}
\end{align}

\begin{algorithm}
\textbf{Step 1: Proposed Fast Channel Tracking Algorithm}\\
\smallskip
\textbf{Input}: $~d$, $~ N$, $~ \lambda$, $~ \psi_{n}$\\
\textbf{Output}: $~\mathbf{\theta}_{k}$\\
\textbf{while $E_r < \tilde{E}_r$}\\
\textbf{for} $t \leqslant 3$\\
\textbf{Conventional channel estimation in the first 3 timeslots}\
\begin{enumerate}
  \item Estimate the beamspace channel 
  \item Find the position of the strongest element $n_{i,k}$ of $\tilde{\mathbf{h}}_{i,k}(t)$.
  \item Approximate the AoA as in (\ref{Eq:theta_approx})
  \setcounter{a}{\value{enumi}}
  \end{enumerate}
\textbf{end for}\\
\textbf{end while}\\

\textbf{for} $t > 3$\\
\textbf{Position prediction}\
\begin{enumerate}
  \setcounter{enumi}{\value{a}}
  \item Predict according to (\ref{Eq:pos_pred}) and (\ref{Eq:theta_pred})
  \item Detect $\mathrm{supp}(\tilde{\mathbf{h}}_k)$ according to (\ref{Eq:supp})
  \item Estimate the nonzero elements of $\tilde{h}_k$
  \item Refine $\theta_{i,k}(t)$ based on $n_{i,k}$ as in (\ref{Eq:theta_approx})
  \setcounter{a}{\value{enumi}}
\end{enumerate}
\textbf{end for}\\

\textbf{Step 2: Ranging}
\begin{enumerate}
  \setcounter{enumi}{\value{a}}
  \item Estimate the BS-UE distance as in (\ref{Eq:dist})\\
  \setcounter{a}{\value{enumi}}
\end{enumerate}

\textbf{Step 3: Localization}
\begin{enumerate}
  \setcounter{enumi}{\value{a}}
  \item Calculate $\theta_{un_{i,k}}(t)$ as in (\ref{eq:th_uni})
  \item Calculate the position of the UE using the estimated AoA and distance as in (\ref{eq:pos})
  \setcounter{a}{\value{enumi}}\\
\end{enumerate}

\textbf{Step 4: Cooperation}
\begin{enumerate}
  \setcounter{enumi}{\value{a}}
  \item Calculate the mean $\overline{r}_k$ as in (\ref{Eq:pos_mean}).
  \item Calculate $~\theta_{i,k}(t)$ from the position of the UE as in (\ref{eq:th_af_pos}).
\end{enumerate}
\caption{Proposed Algorithm}
\end{algorithm}
%The prediction helps track the UE, with low overhead.

\section{Simulation results \& discussions}
\begin{figure}
\centering
%\captionsetup{justification=centering}
\includegraphics[width=0.7\linewidth,trim=0 0 0 0,clip=false]{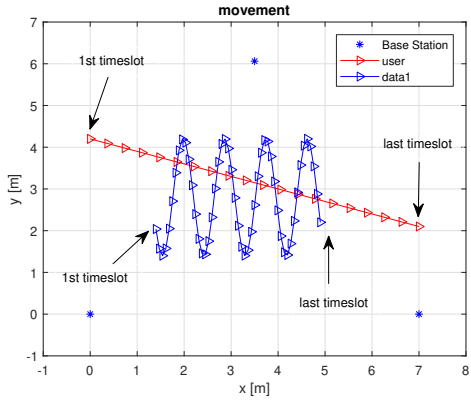}
\caption{BS and UE position}
\label{fig:bs_ue_pos}
\vspace{-0.3cm}
\end{figure}

In this section, we validate the effectiveness of the proposed approach with Monte-Carlo simulations.  In more detail, we assume the following insightful scenario. As illustrated in Fig. \ref{fig:bs_ue_pos}, the BSs are placed on the vertices of equilateral triangle, with the distance between them being 50 m. The use of multiple BSs, means that there are multiple estimations of the UE's position and each BS estimates a different channel. Furthermore, even if the prediction is not very accurate, the result differs for every BS. The beamspace channel is considered a sparse vector with sparsity $V=16$.\\
\indent The evaluation of the algorithms is done using deafness and the probability of successful AoA estimation as measures. Deafness is the power leakage caused by the estimation error and is defined as the estimation error normalized to half the beamwidth and depicted as a percentage. If it reaches 100\%, the estimation has failed as the UE is outside the beam and the algorithm has to start over. The failed estimations are not shown in the figures, in order to show when the algorithm has to start over. The probability is 100\%, if the estimations are always within the half of the beamwidth. Both the FCT and the proposed algorithm, use 128 pilots for channel estimation, in the 3 first timeslots and 16 pilots thereafter. If they fail to estimate correctly, they start over and use 128 pilots for channel estimation. We consider two types of motions. The first one is linear and the other a sinusoidal. In both motions, the average speed, is set to 10 km/h, which, although, is very high for indoor environments, it allows us to evaluate the accuracy of the prediction of each method and the advantages of using multiple BSs to track the UE and their cooperation, under worst case scenario. The frequency of the estimations per second is set to 1. The pathloss is evaluated according to the propagation model presented in \cite{Kokkoniemi2018}. Moreover, it is assumed that the UE employs an omnidirectional antenna, while each BS employs a DLA antenna with 256 elements, each with antenna spacing $d=\dfrac{\lambda}{2}$. The orientation of the first and third BS's antennas are $\dfrac{\mathbf{\pi}}{2}$, while for the second one is $-\dfrac{\mathbf{\pi}}{2}$. The transmitted power, frequency and bandwidth of the UE are 10 dBm, 275 GHz and 40 MHz, respectively. Finally, we assume standard environmental conditions, temperature $T= 296$ K, air pressure $p= 101325$ Pa and relative humidity $\phi= 50$ \%.

\begin{figure}
    \centering
    %\captionsetup{justification=centering,margin=2cm}
    %\begin{minipage}[l]{2.0\columnwidth}
    %\begin{center}
   % \begin{subfigure}[t]{0.3\textwidth}
        \includegraphics[width=0.3\textwidth]{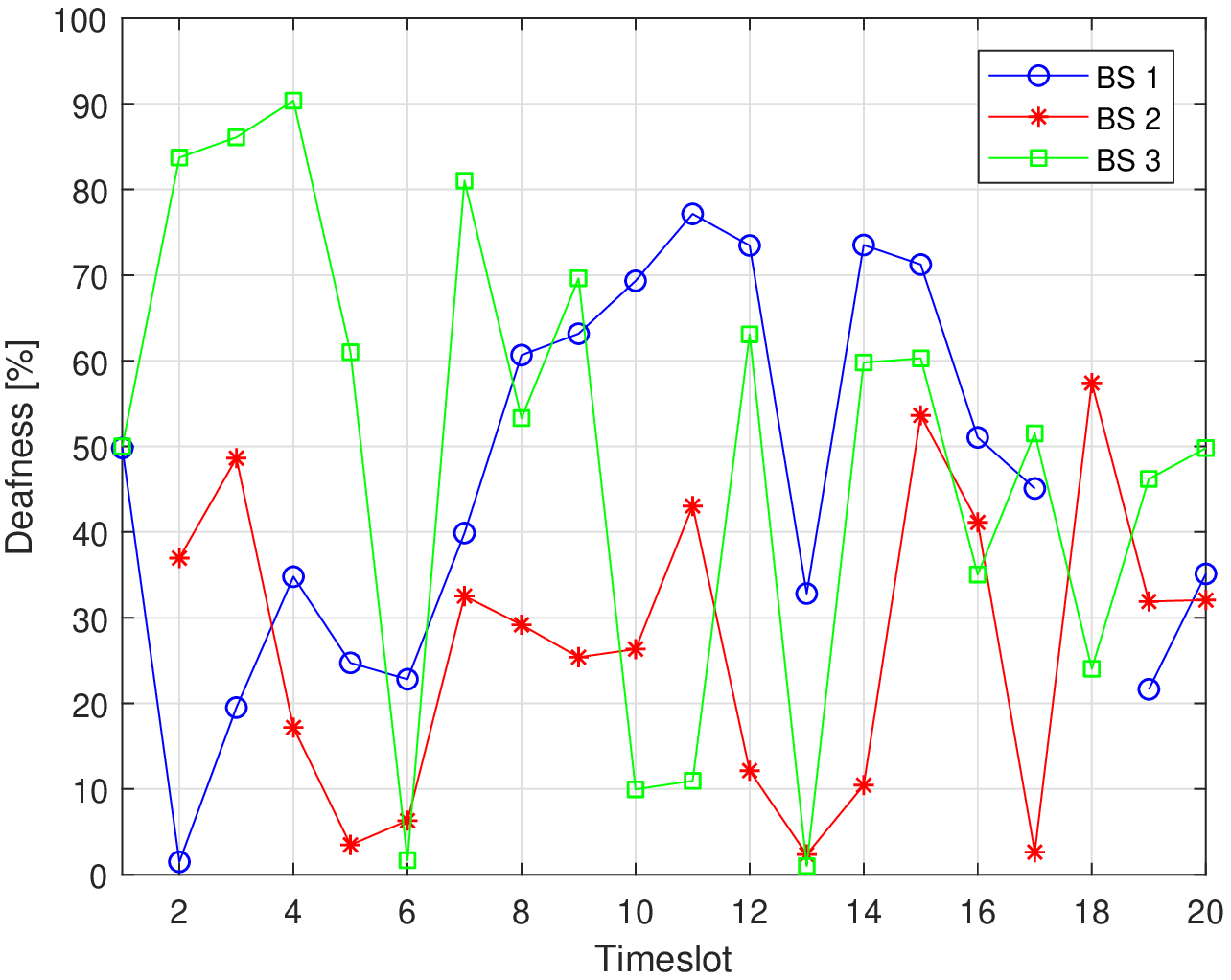}\\
        (a)\\
        %\label{sf:user1_d_fct}
   % \end{subfigure}%
    ~ %add desired spacing between images, e. g. ~, \quad, \qquad etc.
    %\begin{subfigure}[t]{0.3\textwidth}
        \includegraphics[width=0.3\textwidth]{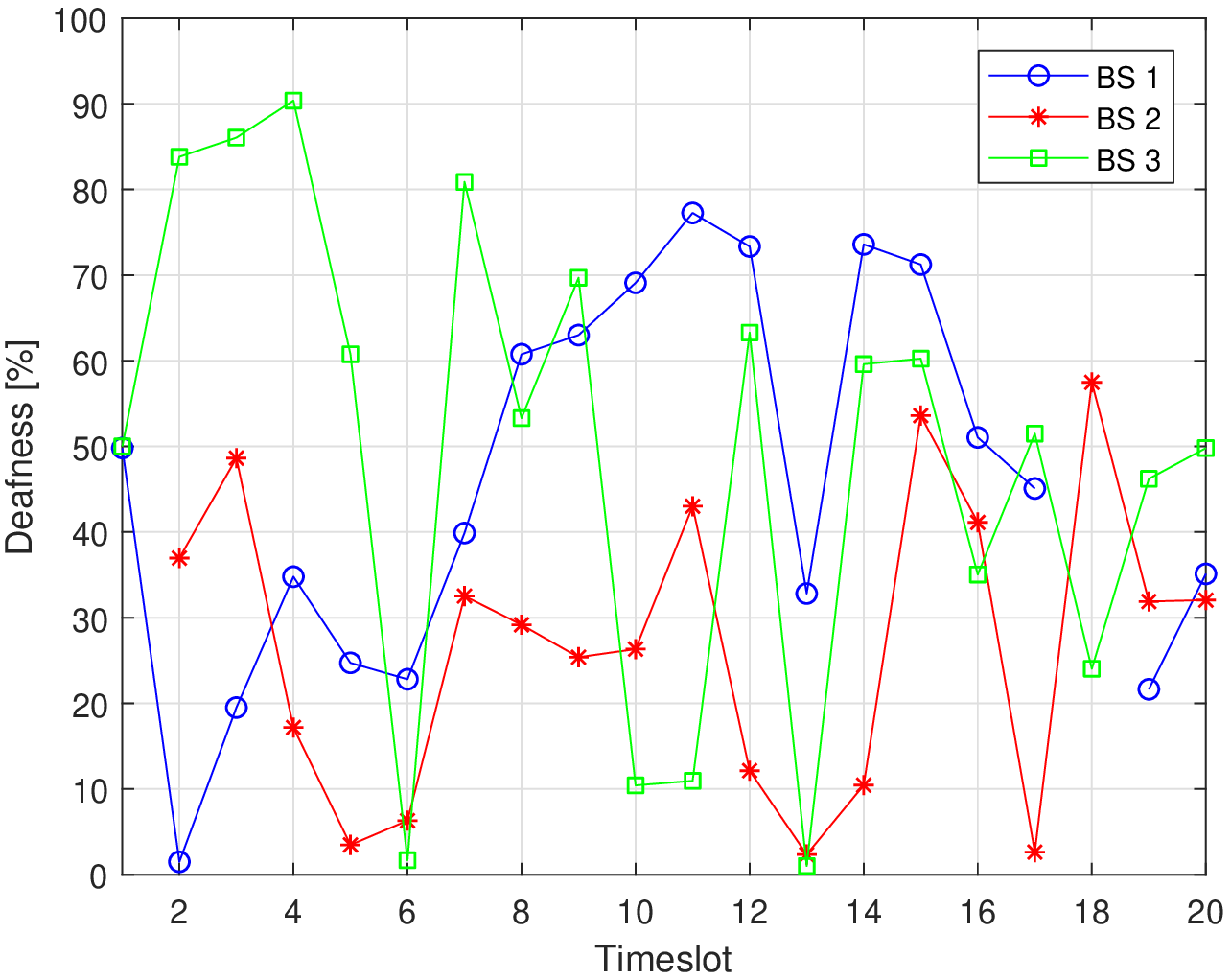}\\
        (b)\\
       % \caption{}
        %\label{sf:user1_d_bL}
    %\end{subfigure}%
    %\begin{subfigure}[t]{0.3\textwidth}
        \includegraphics[width=0.3\textwidth]{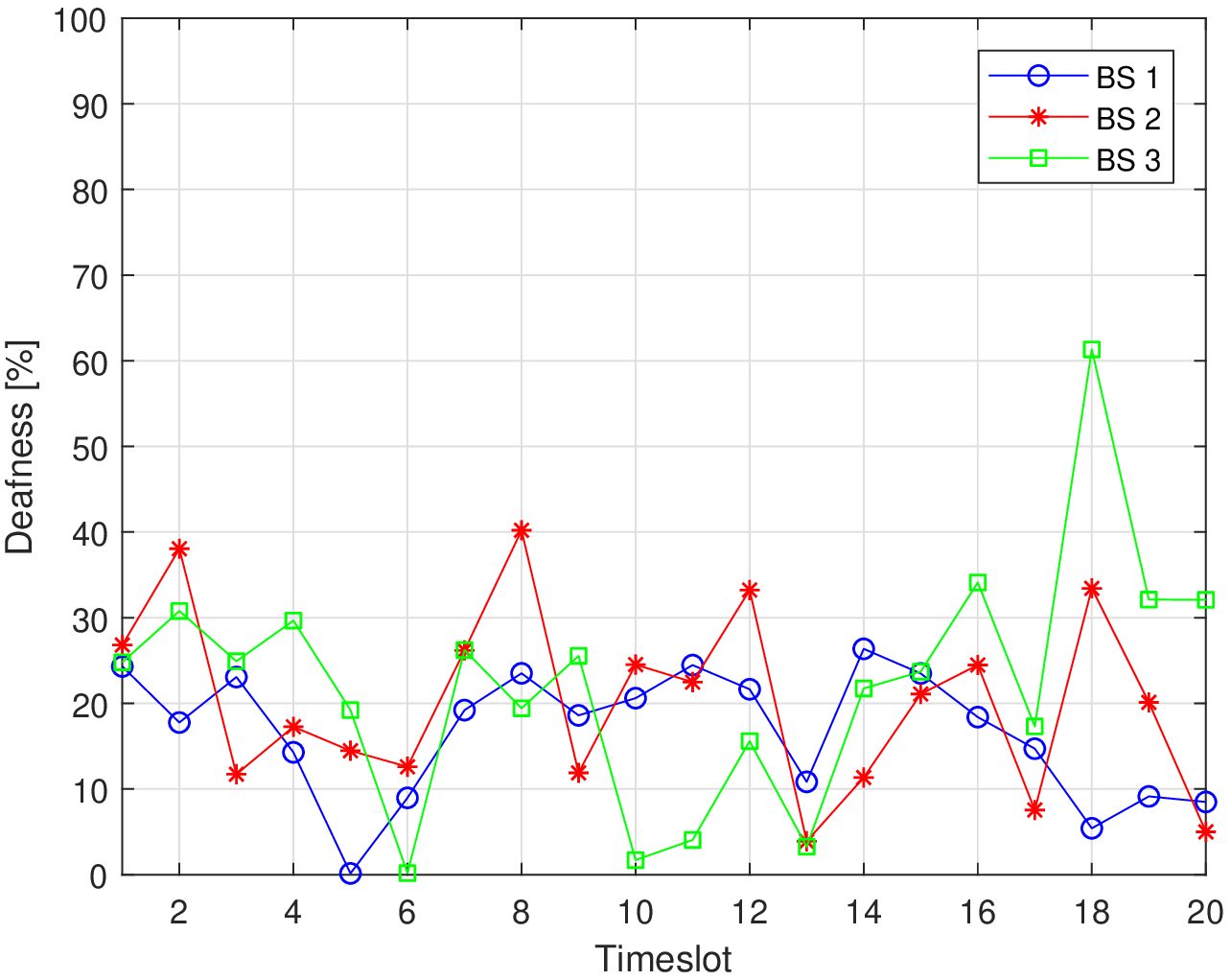}\\
        (c)\\
        %\caption{}
        %\label{sf:user1_d_aL}
    %\end{subfigure}%
    \caption{Linear motion: Average deafness per timeslot in, (a) FCT, (b) Proposed algorithm without cooperation and (c) Proposed algorithm with cooperation.}\label{fig:u1_deaf}
    %\end{center}
   % \end{minipage}
    ~ %add desired spacing between images, e. g. ~, \quad, \qquad etc.
\end{figure}

\indent Fig. \ref{fig:u1_deaf} compares the accuracy of the proposed approach against the FTC algorithm, for each timeslot of the linear motion. From this figure, we observe that both the FCT and the proposed algorithm without the BSs cooperation, share the same level of deafness. From this figure, it is evident that BS 1 cannot find the UE in the 18-th timeslot and BS 2 cannot find it in the first timeslot. This is the result of using the beamspace channel to estimate the AoA. As explained in Phase 1, the tracking algorithm estimates specific AoAs, regardless of the actual one. In this case, the AoAs estimated by BS 1 and 2 result in an error that is higher than the half of the beamwidth. The cooperation results in reducing the overall average deafness to the level of the most accurate BS, from 40\% (in FCT) to 20\%, while also fixing the aforementioned problem. In this case the most accurate BS is the second one, as its position, relative to the position of the UE in each timeslot, results in small changes in the direction that the UE is at. It can be observed, that although the decision of the position is common after phase 4, the result is a different AoA estimation for each BS, due to their position relative to the UE's, which results in 3 different deafness events. The same happens with the probability of successful AoA estimation.%\\

\begin{figure}
    \centering
   % \captionsetup{justification=centering,margin=2cm}
    %\begin{minipage}[l]{2.0\columnwidth}
    %\begin{center}
   % \begin{subfigure}[t]{0.3\textwidth}
        \includegraphics[width=0.3\textwidth]{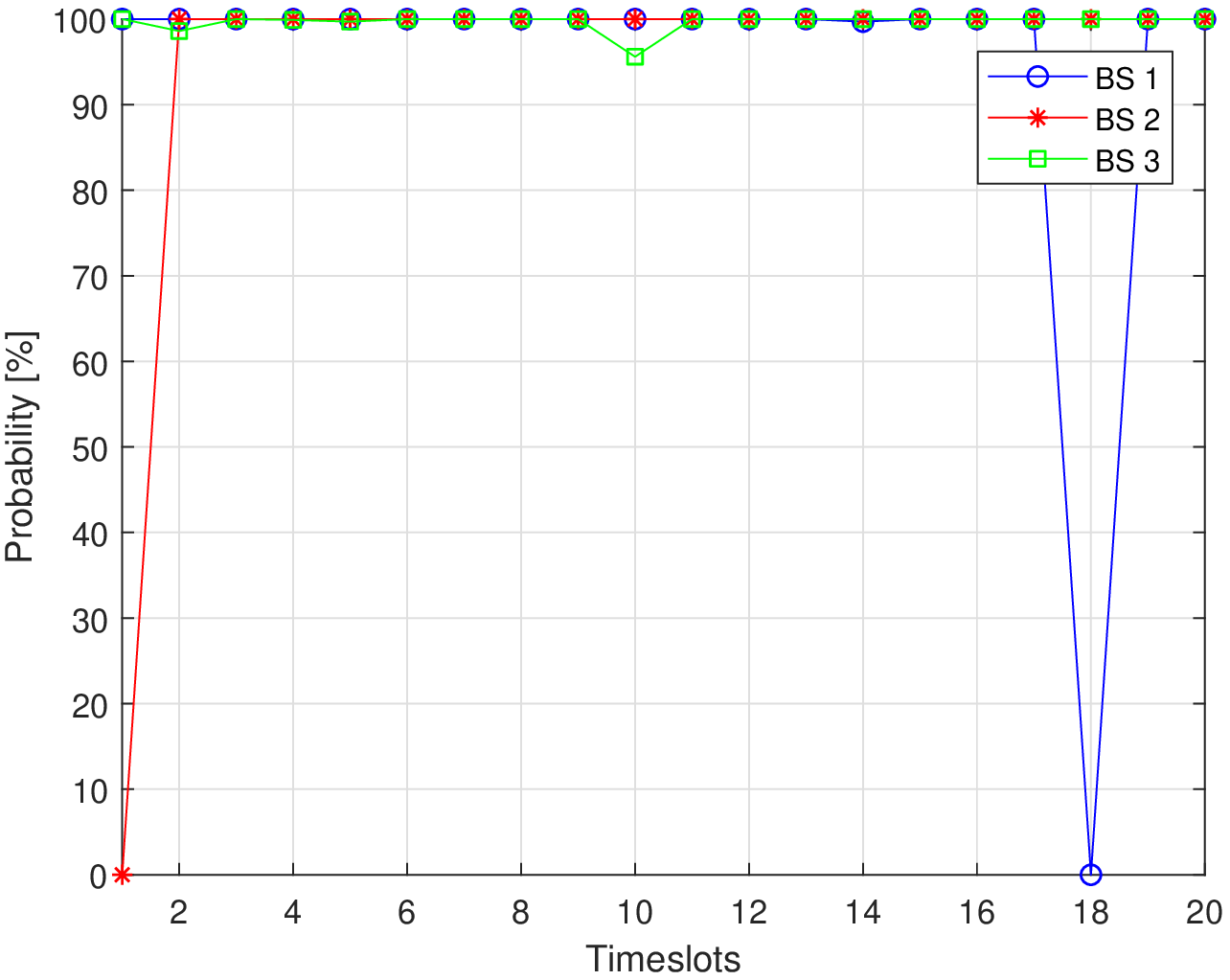}\\
        (a) \\
        %\caption{}
       % \label{sf:user1_p_fct}
    %\end{subfigure}%
    %~ %add desired spacing between images, e. g. ~, \quad, \qquad etc.
    %\begin{subfigure}[t]{0.3\textwidth}
        \includegraphics[width=0.3\textwidth]{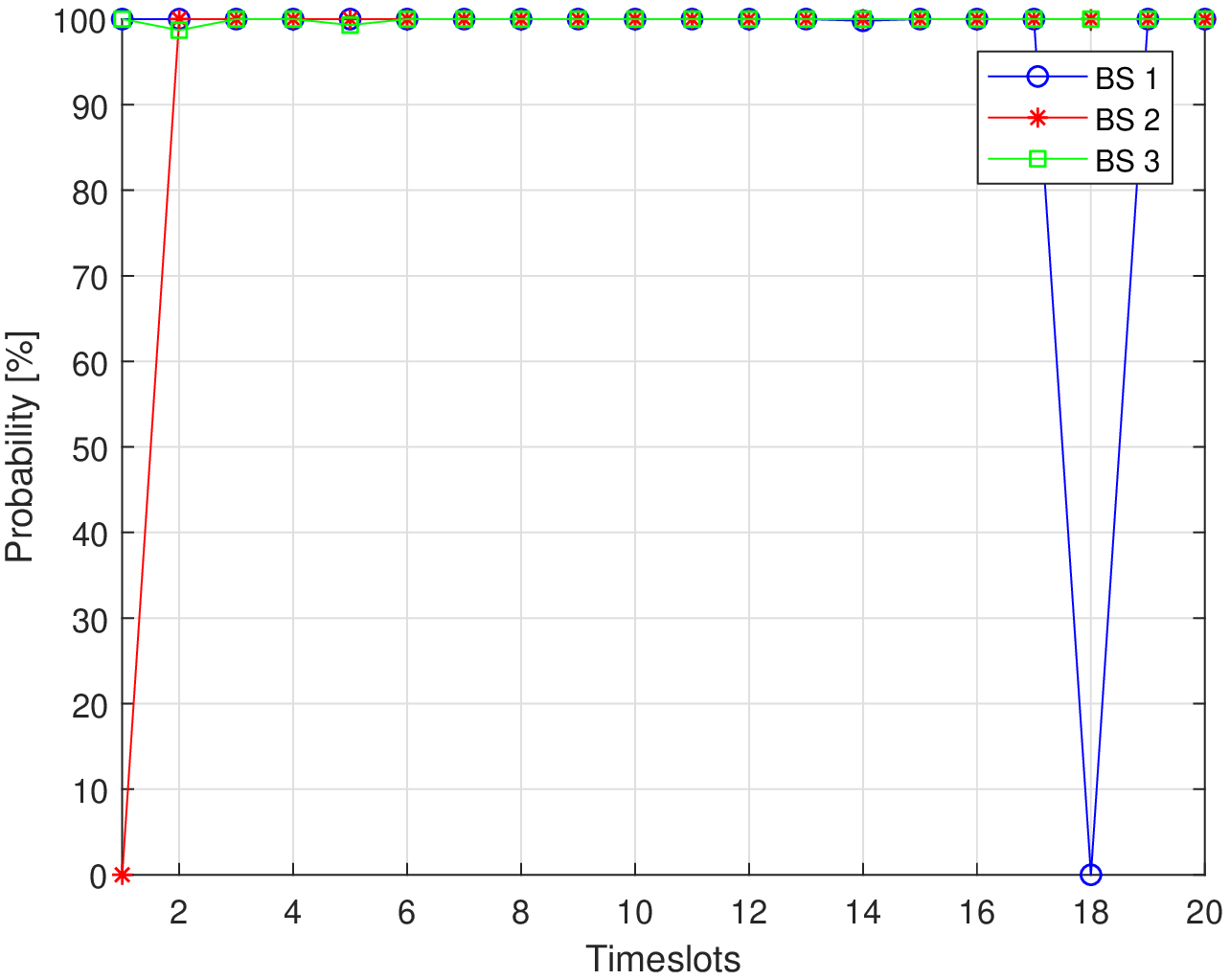}\\
        (b) \\
        %\caption{}
        %\label{sf:user1_p_bL}
    %\end{subfigure}%
    %\begin{subfigure}[t]{0.3\textwidth}
        \includegraphics[width=0.3\textwidth]{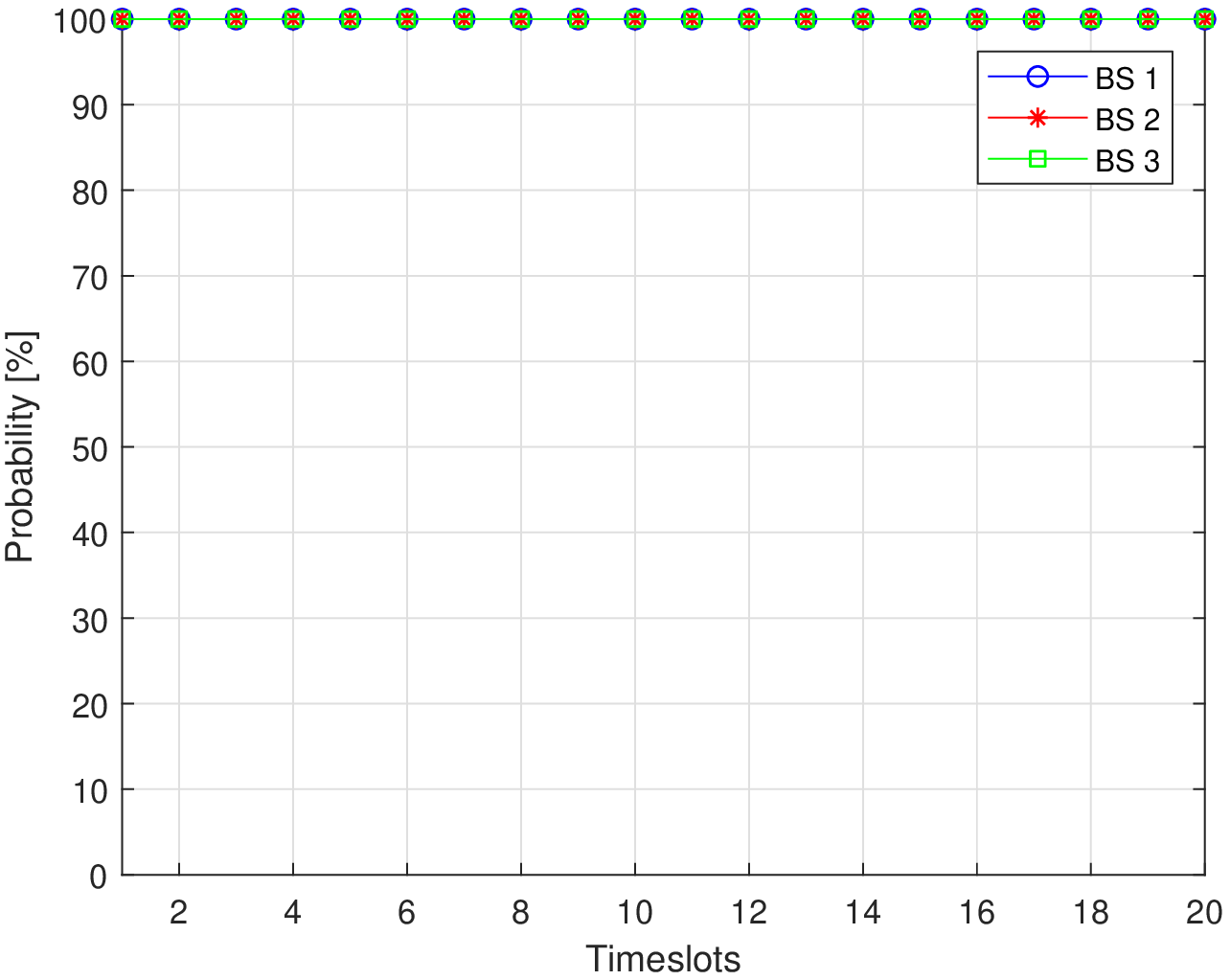}\\
        (c) \\
       % \caption{}
       % \label{sf:user1_p_aL}
%    \end{subfigure}%
    %~ %add desired spacing between images, e. g. ~, \quad, \qquad etc.
    \caption{Linear motion: Probability of successful AoA estimation per timeslot in, (a) FCT, (b) Proposed algorithm without cooperation and (c) Proposed algorithm with cooperation.}\label{fig:u1_prob}
%    \end{center}
    %\end{minipage}
\end{figure}

\indent In Fig. \ref{fig:u1_prob}, the probability of successful AoA estimation per timeslot, is plotted for the linear motion. Both the FCT and the proposed algorithm without the cooperation have the same performance. The BSs also have the same performance, with the exception of the first timeslot for BS 2, the 10-th timeslot for BS3 and the 18-th timeslot for BS 1. The degradation of the first and second BSs was explained previously. The degradation of the third BS in the 10-th timeslot, with the FCT, is caused by the errors in the previous timeslots, along with the low SNR, that cause the prediction to fail. On the other hand, the prediction of the proposed algorithm in this motion is more accurate than the original, as the probability of successful estimation of third BS in the 10-th timeslot is 100\%. The cooperation of the BSs ensures that all the BSs track accurately the direction that the UE is at and can achieve 100\% probability of successful AoA estimation. This is the result of sharing the information of the position of the UE to all the BSs, instead of the BSs operating individually.

\begin{figure}
    \centering
   % \captionsetup{justification=centering,margin=2cm}
    %\begin{minipage}[l]{2.0\columnwidth}
   % \begin{center}
   % \begin{subfigure}[t]{0.3\textwidth}
        \includegraphics[width=0.3\textwidth]{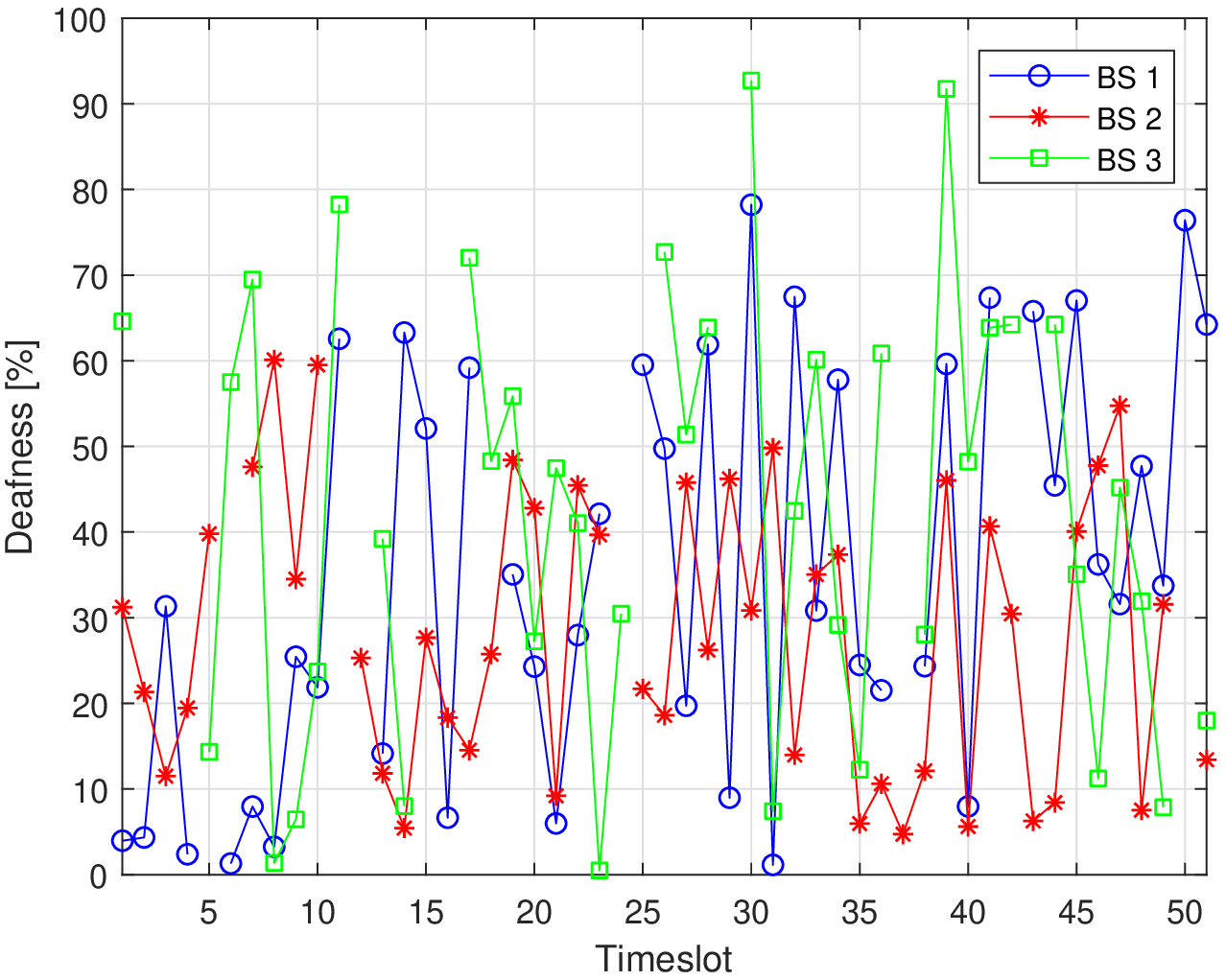}\\
        (a) \\
       % \caption{}
        %\label{sf:user2_d_fct}
   % \end{subfigure}%
   % ~ %add desired spacing between images, e. g. ~, \quad, \qquad etc.
    %\begin{subfigure}[t]{0.3\textwidth}
        \includegraphics[width=0.3\textwidth]{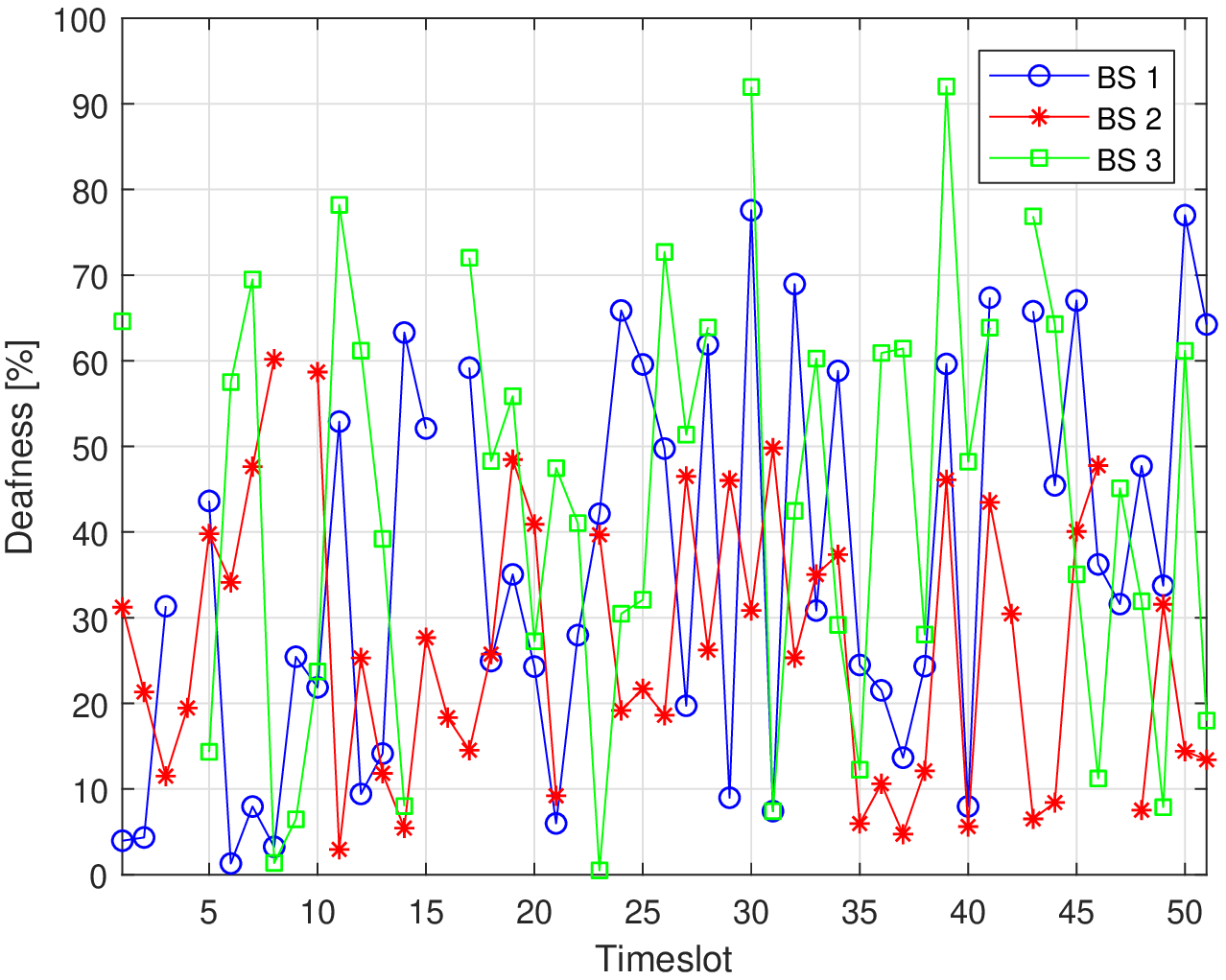}
        \\
        (b) \\
        %\caption{}
        %\label{sf:user2_d_bL}
    %\end{subfigure}%
   % \begin{subfigure}[t]{0.3\textwidth}
        \includegraphics[width=0.3\textwidth]{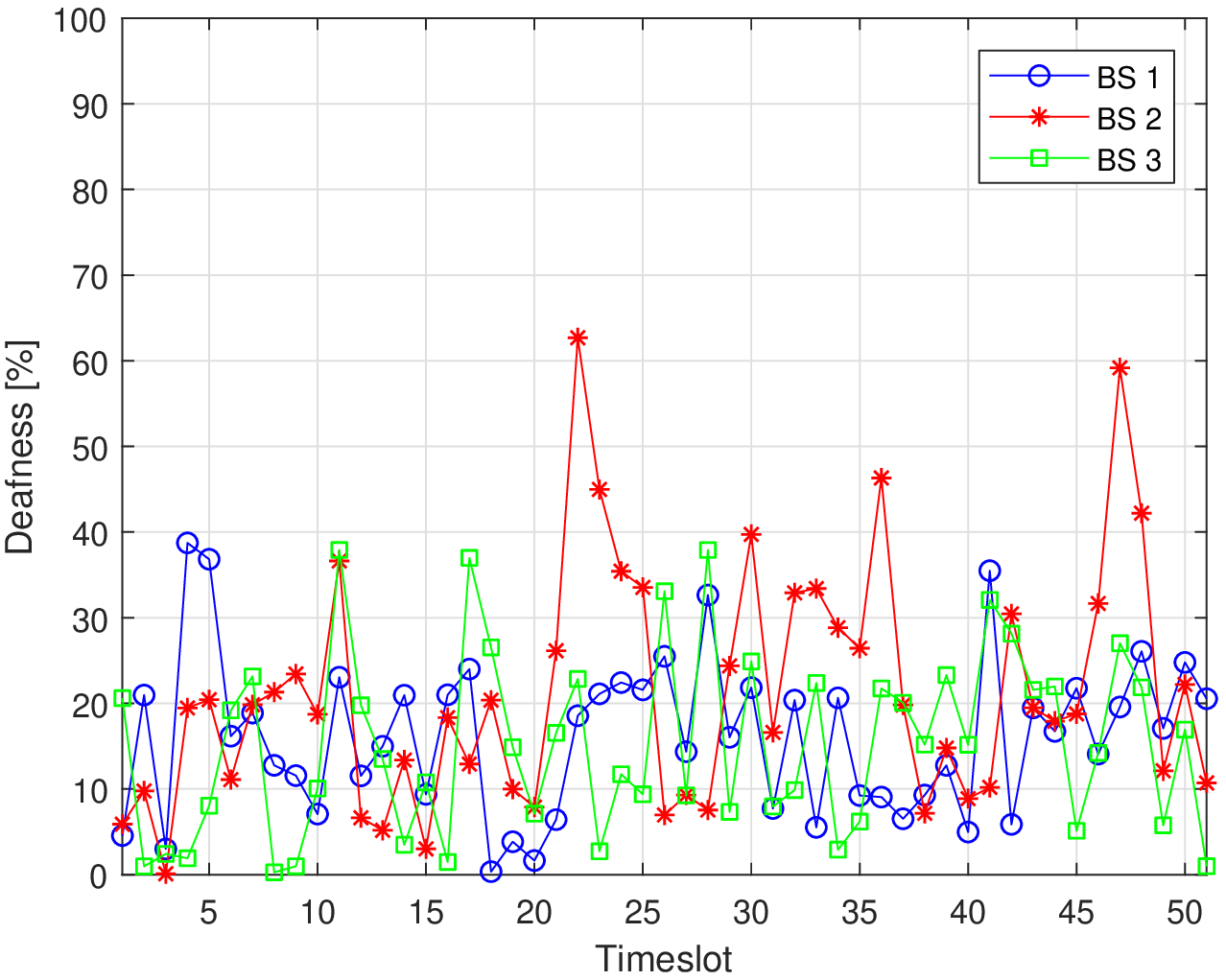}
        \\
        (c) \\
       % \caption{}
       % \label{sf:user2_d_aL}
   % \end{subfigure}%
   % ~ %add desired spacing between images, e. g. ~, \quad, \qquad etc.
    \caption{Sinusoidal motion: Average deafness per timeslot in, (a) FCT, (b) Proposed algorithm without cooperation and (c) Proposed algorithm with cooperation.}\label{fig:u2_deaf}
    %\end{center}
    %\end{minipage}
\end{figure}

\indent Fig. \ref{fig:u2_deaf} depicts the accuracy of the proposed algorithm against the FCT, in each timeslot of the second motion. Both FCT and the proposed algorithm without the cooperation result in the same overall deafness. 
The behavior of deafness is random because as mentioned previously, the FCT algorithm estimates specific directions regardless of the actual AoA of the UE. 
Furthermore, all BSs cannot track the UE consistently, due to the abrupt changes in direction of the second motion. The cooperation helps reduce the deafness from $35\%$, to $17.5\%$, while guaranteeing that all the BSs are able to accurately know the AoA of the UE. 
%In Fig. \ref{fig:u2_pilots}, the FCT and the proposed algorithm more often than not fails to find the direction of the UE, due to the prediction failing and the number of pilots needed for channel estimation increases. In the case of the FCT, every 6 timeslots on average of using 16 pilots for channel estimation, the algorithm fails to estimate correctly. The new prediction of the proposed algorithm seems to work a little better, but not to a satisfactory level. The cooperation helps keep the number of pilots needed to a low level.

\begin{figure}
    \centering
   % \captionsetup{justification=centering,margin=2cm}
    %\begin{minipage}[l]{2.0\columnwidth}
   % \begin{center}
   % \begin{subfigure}[t]{0.3\textwidth}
        \includegraphics[width=0.3\textwidth]{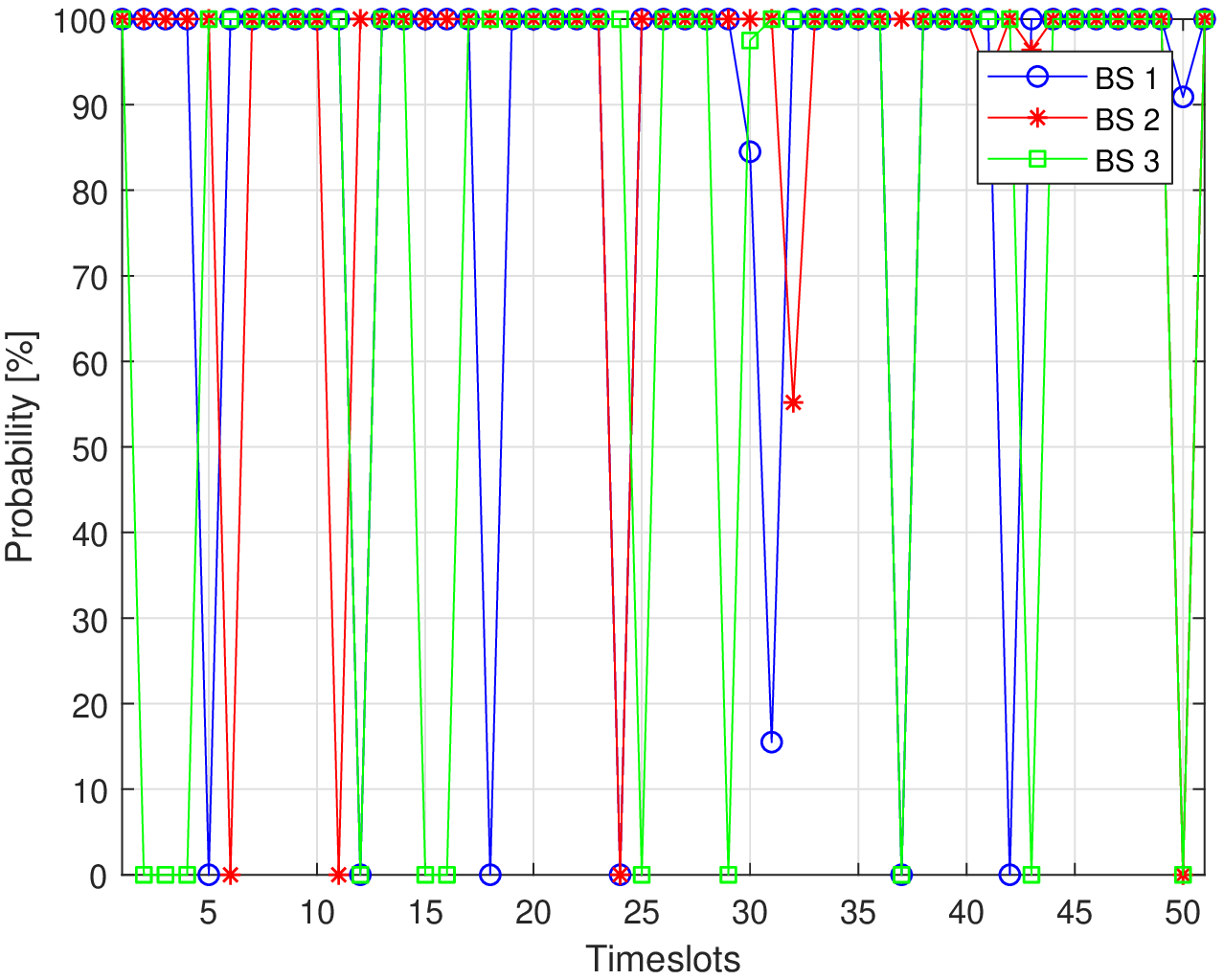}
        \\ (a) \\
        %\caption{}
       % \label{sf:user2_p_fct}
   % \end{subfigure}%
   % ~ %add desired spacing between images, e. g. ~, \quad, \qquad etc.
   % \begin{subfigure}[t]{0.3\textwidth}
        \includegraphics[width=0.3\textwidth]{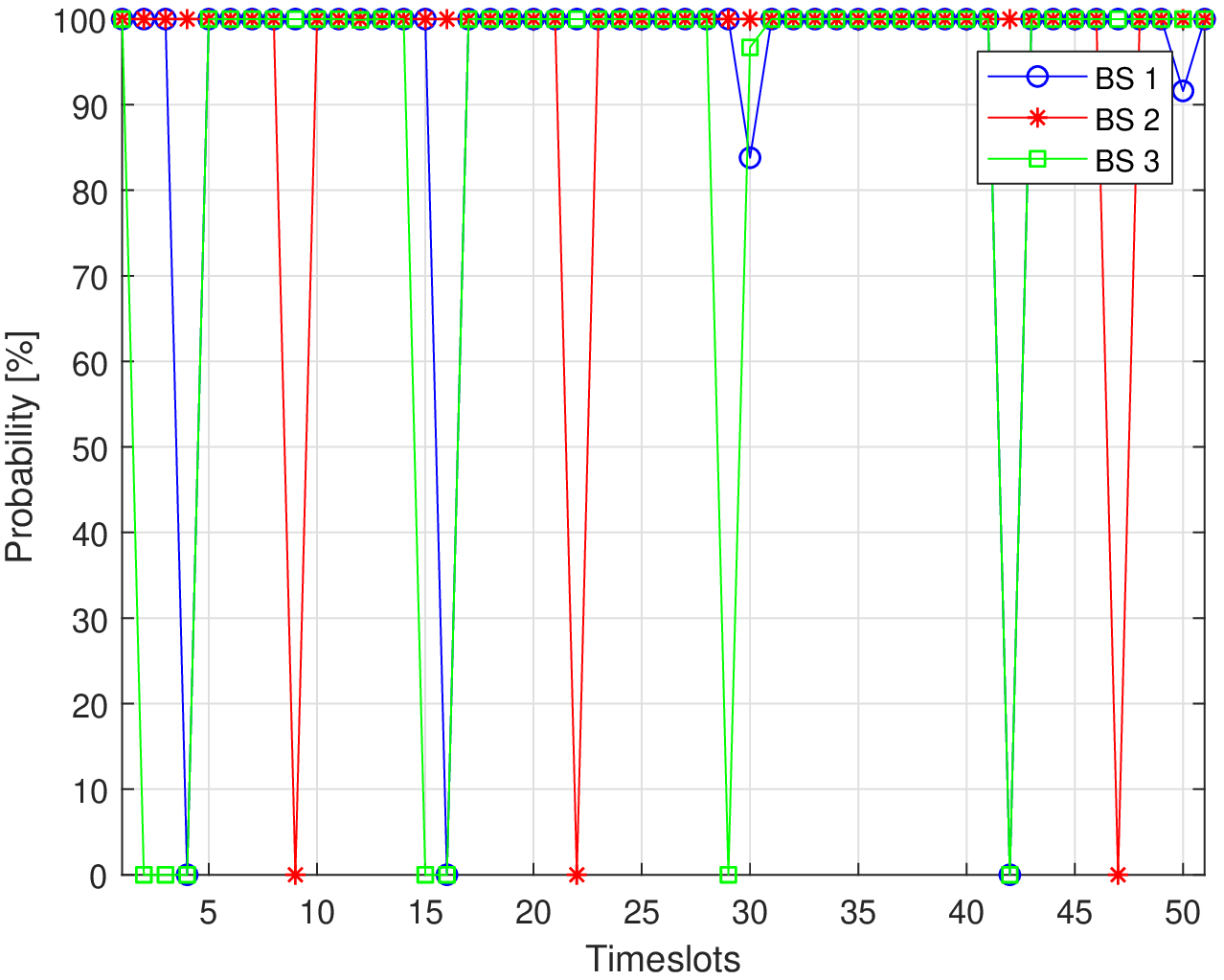}
        \\ (b) \\
       % \caption{}
       % \label{sf:user2_p_bL}
   % \end{subfigure}%
   % \begin{subfigure}[t]{0.3\textwidth}
        \includegraphics[width=0.3\textwidth]{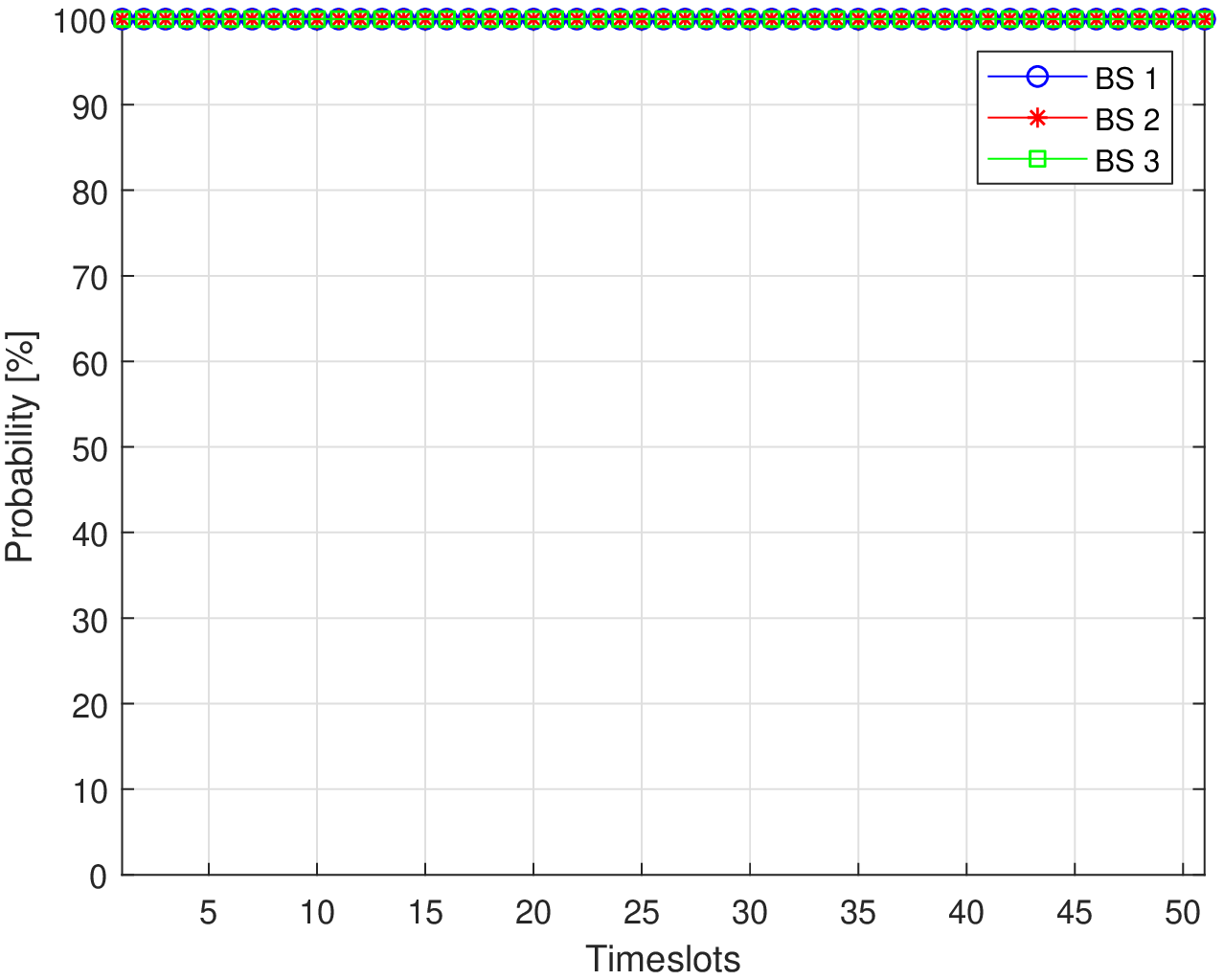}
        \\ (c) \\
        %\caption{}
       % \label{sf:user2_p_aL}
   % \end{subfigure}%
   % ~ %add desired spacing between images, e. g. ~, \quad, \qquad etc.
    \caption{Sinusoidal motion: Probability of successful AoA estimation  per timeslot in, (a) FCT, (b) Proposed algorithm without cooperation and (c) Proposed algorithm with cooperation.}\label{fig:u2_prob}
   % \end{center}
    %\end{minipage}
\end{figure}

\indent Fig. \ref{fig:u2_prob} illustrates the probability of estimate  AoA estimation per timeslot for the sinusoidal motion. From this figure, it is evident that both the proposed algorithm  without cooperation and the FCT fail most of the time to find the AoA of the UE. This is the result of the failure of the predictions to point at the right direction. From this figure, it can be observed that the proposed algorithm without cooperation, outperforms the FCT; however since in both algorithms the BSs operate independetly, their performance is not satisfactory. The cooperation achieves 100\% probability of successful AoA estimation, as at least one BS in each timeslot finds the AoA of the UE. This is the result of the position and orientation of the BSs, as they allow each of them to estimate a different AoA and is the reason why the cooperation can achieve this probability.

%\clearpage

\section{Conclusion}
In this paper, we provided a novel cooperation aided localization approach for indoor THz communication systems that, although it requires low overhead, it provides high estimation accuracy and countermeasures the deafness problem. The efficiency of the approach was validated by respective simulation results that reveal that the new prediction behaves a little better than the original in both linear motions and motions with abrupt changes in direction and the cooperation scheme reduces the deafness by half, while also guaranteeing 100\% probability of successfull AoA estimation.

\balance
\bibliographystyle{IEEEtran}
\bibliography{IEEEabrv,ReferencesLAT}

\begin{thebibliography}{10}
\providecommand{\url}[1]{#1}
\csname url@samestyle\endcsname
\providecommand{\newblock}{\relax}
\providecommand{\bibinfo}[2]{#2}
\providecommand{\BIBentrySTDinterwordspacing}{\spaceskip=0pt\relax}
\providecommand{\BIBentryALTinterwordstretchfactor}{4}
\providecommand{\BIBentryALTinterwordspacing}{\spaceskip=\fontdimen2\font plus
\BIBentryALTinterwordstretchfactor\fontdimen3\font minus
  \fontdimen4\font\relax}
\providecommand{\BIBforeignlanguage}[2]{{%
\expandafter\ifx\csname l@#1\endcsname\relax
\typeout{** WARNING: IEEEtran.bst: No hyphenation pattern has been}%
\typeout{** loaded for the language `#1'. Using the pattern for}%
\typeout{** the default language instead.}%
\else
\language=\csname l@#1\endcsname
\fi
#2}}
\providecommand{\BIBdecl}{\relax}
\BIBdecl

\bibitem{PhD:Boulogeorgos}
A.-A.~A. Boulogeorgos, ``Interference mitigation techniques in modern wireless
  communication systems,'' Ph.D. dissertation, Aristotle University of
  Thessaloniki, Thessaloniki, Greece, Sep. 2016.

\bibitem{A:LC_CR_vs_SS}
A.-A.~A. Boulogeorgos and G.~K. Karagiannidis, ``Low-cost cognitive radios
  against spectrum scarcity,'' \emph{IEEE Technical Committee on Cognitive
  Networks Newsletter}, vol.~3, no.~2, pp. 30--34, Nov. 2017.

\bibitem{A:Analytical_Performance_Assessment_of_THz_Wireless_Systems}
A.-A.~A. Boulogeorgos, E.~N. Papasotiriou, and A.~Alexiou, ``Analytical
  performance assessment of {THz} wireless systems,'' \emph{IEEE Access},
  vol.~7, no.~1, pp. 1--18, Jan. 2019.

\bibitem{Jornet2011}
J.~M. Jornet and I.~F. Akyildiz, ``Channel modeling and capacity analysis for
  electromagnetic wireless nanonetworks in the terahertz band,'' \emph{{IEEE}
  Transactions on Wireless Communications}, vol.~10, no.~10, pp. 3211--3221,
  Oct. 2011.

\bibitem{our_PIMRC}
E.~N. Papasotiriou, J.~Kokkoniemi, J.~L. A.-A. A.~Boulogeorgos, A.~Alexiou, and
  M.~Juntti, ``A new look to 275 to 400 {GHz} band: {Channel} model and
  performance evaluation,'' in \emph{IEEE International Symposium on Personal,
  Indoor and Mobile Radio Communications (PIMRC)}, Bolonia, Italy, Sep. 2018.

\bibitem{Kokkoniemi2018}
K.~J. Lehtomaki and M.~Juntti, ``Simplified molecular absorption loss model for
  275 – 400 gigahertz frequency band,'' \emph{Proc. European Conf. Antennas
  Propag.}, Jan. 2018.

\bibitem{C:PerfEvaluation}
A.-A.~A. Boulogeorgos, E.~N. Papasotiriou, J.~Kokkoniemi, J.~Lehtom\"aki,
  A.~Alexiou, and M.~Juntti, ``Performance evaluation of {THz} wireless systems
  operating in 275-400 {GHz} band,'' in \emph{IEEE 87th Vehicular Technology
  Conference: International Workshop on THz Communication Technologies for
  Systems Beyond 5G}, Porto, Portugal, Jun. 2018.

\bibitem{our_spawc_paper_2018}
A.-A.~A. Boulogeorgos, E.~N. Papasotiriou, and A.~Alexiou, ``A distance and
  bandwidth dependent adaptive modulation scheme for {THz} communications,'' in
  \emph{19th IEEE International Workshop on Signal Processing Advances in
  Wireless Communications (SPAWC)}, Kalamata, Greece, Jul. 2018.

\bibitem{C:UserAssociationInUltraDenseTHzNetworks}
A.-A.~A. Boulogeorgos, S.~Goudos, and A.~Alexiou, ``Users association in ultra
  dense {THz} networks,'' in \emph{IEEE International Workshop on Signal
  Processing Advances in Wireless Communications (SPAWC)}, Kalamata, Greece,
  Jun. 2018.

\bibitem{Boulogeorgos2018}
A.-A.~A. Boulogeorgos, A.~Alexiou, T.~Merkle, C.~Schubert, R.~Elschner,
  A.~Katsiotis, P.~Stavrianos, D.~Kritharidis, P.-K. Chartsias, J.~Kokkoniemi,
  M.~Juntti, J.~Lehtomaki, A.~Teixeira, and F.~Rodrigues, ``Terahertz
  technologies to deliver optical network quality of experience in wireless
  systems beyond {5G},'' \emph{{IEEE} Communications Magazine}, vol.~56, no.~6,
  pp. 144--151, Jun. 2018.

\bibitem{WP:Wireless_Thz_system_architecture_for_networks_beyond_5G}
A.-A.~A. Boulogeorgos, A.~Alexiou, D.~Kritharidis, A.~Katsiotis, G.~Ntouni,
  J.~Kokkoniemi, J.~Lethtomaki, M.~Juntti, D.~Yankova, A.~Mokhtar, J.-C. Point,
  J.~Machodo, R.~Elschner, C.~Schubert, T.~Merkle, R.~Ferreira, F.~Rodrigues,
  and J.~Lima, ``Wireless terahertz system architectures for networks beyond
  {5G},'' TERRANOVA CONSORTIUM, White paper 1.0, Jul. 2018.

\bibitem{Gao2017}
X.~Gao, L.~Dai, Y.~Zhang, T.~Xie, X.~Dai, and Z.~Wang, ``Fast channel tracking
  for terahertz beamspace massive {MIMO} systems,'' \emph{{IEEE} Transactions
  on Vehicular Technology}, vol.~66, no.~7, pp. 5689--5696, Jul. 2017.

\bibitem{Zhang2016}
C.~Zhang, D.~Guo, and P.~Fan, ``Tracking angles of departure and arrival in a
  mobile millimeter wave channel,'' in \emph{{IEEE} International Conference on
  Communications ({ICC})}, May 2016.

\bibitem{Va2016}
V.~Va, H.~Vikalo, and R.~W. Heath, ``Beam tracking for mobile millimeter wave
  communication systems,'' in \emph{{IEEE} Global Conference on Signal and
  Information Processing ({GlobalSIP})}, Dec. 2016.

\bibitem{Jayaprakasam2017}
S.~Jayaprakasam, X.~Ma, J.~W. Choi, and S.~Kim, ``Robust beam-tracking for
  {mmWave} mobile communications,'' \emph{{IEEE} Communications Letters},
  vol.~21, no.~12, pp. 2654--2657, Dec. 2017.

\bibitem{Dargie2010}
W.~Dargie and C.~Poellabauer, \emph{Fundamentals of Wireless Sensor
  Networks}.\hskip 1em plus 0.5em minus 0.4em\relax John Wiley {\&} Sons, Ltd,
  Jul. 2010.

\bibitem{Rong2006}
R.~P and M.~Sichitiu, ``Angle of arrival localization for wireless sensor
  networks,'' in \emph{3rd Annual {IEEE} Communications Society on Sensor and
  Ad Hoc Communications and Networks}, Sep. 2006.

\bibitem{Drawil2013}
N.~M. Drawil, H.~M. Amar, and O.~A. Basir, ``{GPS} localization accuracy
  classification: A context-based approach,'' \emph{{IEEE} Transactions on
  Intelligent Transportation Systems}, vol.~14, no.~1, pp. 262--273, Mar. 2013.

\bibitem{Wymeersch2009}
H.~Wymeersch, J.~Lien, and M.~Z. Win, ``Cooperative localization in wireless
  networks,'' \emph{Proceedings of the {IEEE}}, vol.~97, no.~2, pp. 427--450,
  Feb. 2009.

\bibitem{Ouyang2010}
R.~Ouyang, A.-S. Wong, and C.-T. Lea, ``Received signal strength-based wireless
  localization via semidefinite programming: Noncooperative and cooperative
  schemes,'' \emph{{IEEE} Transactions on Vehicular Technology}, vol.~59,
  no.~3, pp. 1307--1318, Mar. 2010.

\bibitem{Sayeed2013}
A.~Sayeed and J.~Brady, ``Beamspace {MIMO} for high-dimensional multiuser
  communication at millimeter-wave frequencies,'' in \emph{{IEEE} Global
  Communications Conference ({GLOBECOM})}, Dec. 2013.

\bibitem{Lin2017}
H.~Lin, F.~Gao, S.~Jin, and G.~Y. Li, ``A new view of multi-user hybrid massive
  {MIMO}: Non-orthogonal angle division multiple access,'' \emph{{IEEE} Journal
  on Selected Areas in Communications}, vol.~35, no.~10, pp. 2268--2280, Oct.
  2017.

\bibitem{Bronshtein2015}
I.~Bronshtein, K.~Semendyayev, G.~Musiol, and H.~Mühlig, \emph{Handbook of
  Mathematics}.\hskip 1em plus 0.5em minus 0.4em\relax Springer Berlin
  Heidelberg, 2015.

\bibitem{Chang2018}
Y.-J. Chang, C.-H. Ou, and K.-F. Ssu, ``A cluster analysis-based localization
  scheme for wireless sensor networks using mobile anchor nodes with
  directional antennas,'' in \emph{{IEEE} International Conference on Applied
  System Invention ({ICASI})}, Apr. 2018.

\bibitem{Karl2005}
H.~Karl and A.~Willig, \emph{Protocols and Architectures for Wireless Sensor
  Networks}.\hskip 1em plus 0.5em minus 0.4em\relax John Wiley {\&} Sons, Ltd,
  Apr. 2005.

\end{thebibliography}

\end{document}